\documentclass[twocolumn]{svjour3}  
\smartqed  

\usepackage[dvipdfmx]{color}
\usepackage{graphicx}
\usepackage[T1]{fontenc}
\usepackage{mathptmx}  
\usepackage[scaled]{helvet}  
\usepackage{courier}
\usepackage[subrefformat=parens]{subcaption}

\newcommand{\realset}{\mathbb{R}}

\newcommand{\sgn}{\mathrm{sgn}}
\newcommand{\diag}{\mathrm{diag}}

\newcommand{\sat}{\mathrm{sat}}

\newcommand{\kdelta}{k_{\delta}}
\newcommand{\kxi}{k_{\xi}}

\newcommand{\psid}{\psi^{\mathrm{d}}}
\newcommand{\rd}{r^{\mathrm{d}}}
\newcommand{\deltati}{\widetilde{\delta}}
\newcommand{\xiti}{\widetilde{\xi}}

\newcommand{\gdelta}{g_{\delta}}
\newcommand{\fxi}{f_{\xi}}
\newcommand{\gxi}{g_{\xi}}

\newcommand{\alphadelta}{\alpha_{\delta}}
\newcommand{\alphaeta}{\alpha_{\eta}}

\newcommand{\epsi}{e_{\psi}}
\newcommand{\er}{e_{r}}

\newcommand{\markupdraft}[2]{
    \ifthenelse{\equal{#1}{display}}{#2}{}
    \ifthenelse{\equal{#1}{color}}{\color{#2}}{}
}

\makeatletter
\let\cl@chapter\undefined
\makeatletter
\usepackage{amsmath,amssymb}   
\usepackage{mathtools} 
\usepackage{cleveref}  
\Crefname{equation}{Eq.}{Eqs.}%
\Crefname{figure}{Fig.}{Figs}
\Crefname{table}{Tab.}{Tabs.}
\Crefname{section}{Sec.}{Secs.}%
\crefname{equation}{Eq.}{Eqs.}%
\crefname{figure}{Fig.}{Figs.}%
\crefname{table}{Tab.}{Tabs.}
\crefname{section}{Sec.}{Secs.}%
\usepackage{amsbsy}
\usepackage{bm}
\usepackage{algorithmic}
\usepackage{algorithm}

\usepackage{ulem}

\newcommand\Erase{\bgroup\markoverwith{\textcolor{red}{\rule[.5ex]{2pt}{0.4pt}}}\ULon}

\journalname{Journal of Marine Science and Technology}
\begin{document}

\title{
    Nonlinear steering control under input magnitude and rate constraints with exponential convergence
}

\author{
    Rin Suyama
    \and Satoshi Satoh
    \and Atsuo Maki
}

\institute{R. Suyama \and S. Satoh \and A. Maki \at
              Osaka University, 2-1 Yamadaoka, Suita, Osaka, Japan \\
              \email{suyama\_rin@naoe.eng.osaka-u.ac.jp}  \\
              \email{satoh@mech.eng.osaka-u.ac.jp}  \\
              \email{maki@naoe.eng.osaka-u.ac.jp} 
}

\date{Received: date / Accepted: date}

\maketitle

\begin{abstract}
    A ship steering control is designed for a nonlinear maneuvering model whose rudder manipulation is constrained in both magnitude and rate.
    In our method, the tracking problem of the target heading angle with input constraints is converted into the tracking problem for a strict-feedback system without any input constraints.
    To derive this system, hyperbolic tangent ($\tanh$) function and auxiliary variables are introduced to deal with the input constraints.
    Furthermore, using the feature of the derivative of $\tanh$ function, auxiliary systems are successfully derived in the strict-feedback form.
    The backstepping method is utilized to construct the feedback control law for the resulting cascade system.
    The proposed steering control is verified in numerical experiments, and the result shows that the tracking of the target heading angle is successful using the proposed control law.
    
    \keywords{
        Ship Steering Control
        \and Exponential Stability
        \and Input Magnitude Constraint
        \and Input Rate Constraint
        \and Backstepping
    }

\end{abstract}

\section{Introduction}
    \label{sec:intro}

    Ship is one of the transportation that handles the mass transportation of cargo and passengers, and technology for the safe navigation of ships is an important research issue.
    In many cases, ships navigating the oceans utilize steering control laws.
    In the case the target heading angle is a time-invariant constant, the control laws are often referred to as course keeping control \cite{Tani1952,VanAmerongen1982}.

    Response models of ship maneuvering motion and steering controls based on them have long been studied.
    The research on steering control started with the study using Proportional-Integral control in \cite{Minorsky1922}.
    Proportional-Derivative control in \cite{Schiff1949} is also well-known.
    Nomoto's study \cite{Nomoto1957} was the first to consider such a ship course control from the system control point of view.
    In this study, the maneuvering model of a ship was represented as a first-order or second-order system.
    In particular, the first-order model is widely used as the \textit{Nomoto's KT model}, for instance, to evaluate the maneuverability of new ships in ship building companies.
    In the literature \cite{VanAmerongen1982}, the steering control was designed using the model reference adaptive control technique.
    In the literature \cite{McGookin2000}, sliding mode control (SMC) was adopted in the design of the steering control, and the design parameters included in the designed control law were optimized based on the genetic algorithm.
    In the literature \cite{Du2007}, the steering control for a maneuvering model with time-varying uncertain parameters, including control coefficient, was designed using the adaptive backstepping method.
    
    \begin{figure*}[tb]
        \centering
        \includegraphics[width=1.0\hsize]{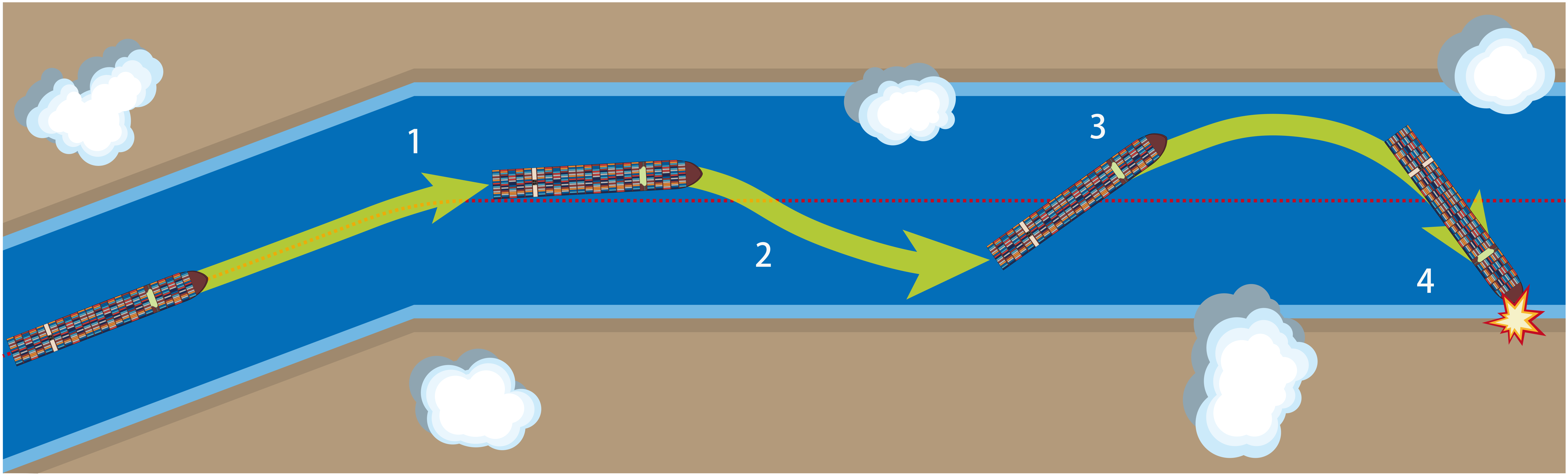}
        \caption{
            An example of the mechanism of allision/collision accidents due to the constraints on rudder manipulation;
            1. The control law sends the rudder command to follow the target path. However the response of maneuvering motion delays due to the constrained manipulation of the steering system.
            2. Tracking error remains, so the control law continues to command the steering system to turn right. As a result, overshoot occurs.
            3. The control law attempts to recover from the overshoot and commands a left turn, but again the constraints on the rudder angle and the steering speed cause a delay in the response of the maneuvering motion.
            4. Overshooting occurs continuously and, in the worst case, is amplified, leading to an allision/collision.
        }
        \label{fig:miss}
    \end{figure*}
    \begin{figure}[tb]
        \centering
        \includegraphics[width=1.0\hsize]{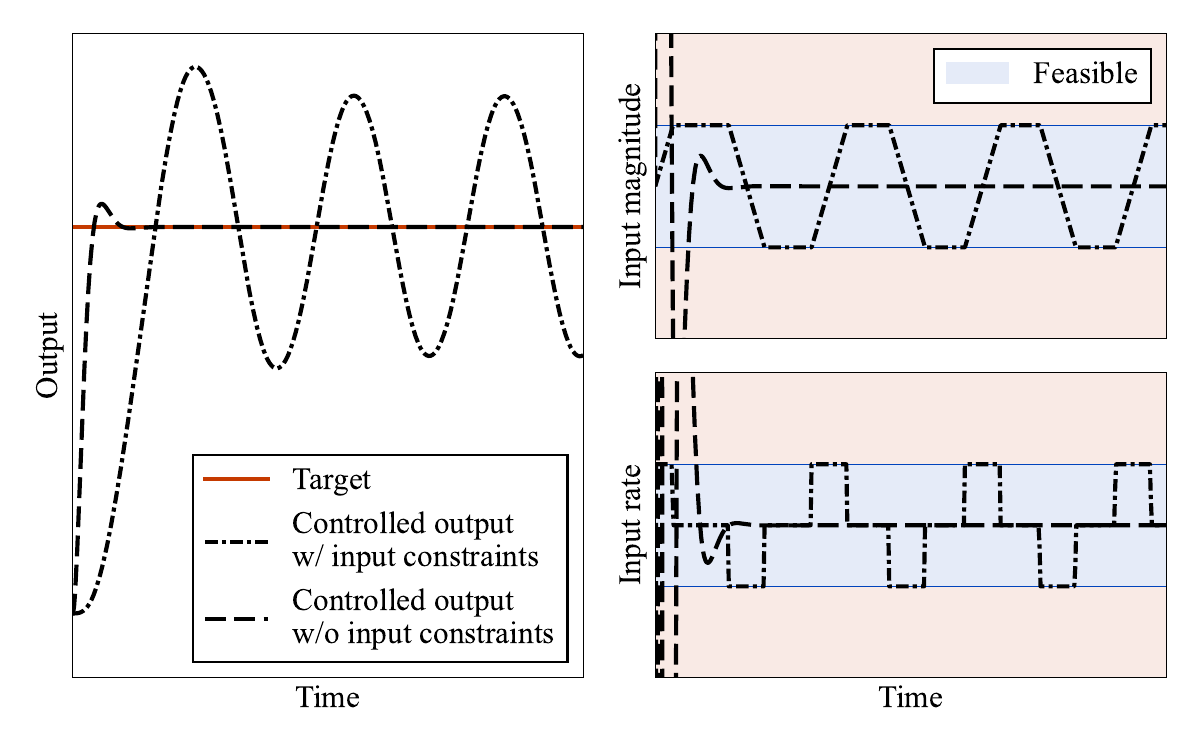}
        \caption{An example of the degradation of a control law due to input magnitude and rate constraints.}
        \label{fig:degradation}
    \end{figure}
    In a ship maneuvering mechanism, there are constraints on the manipulation of the actuators such as rudders and propellers.
    These input constraints are due to the mechanics of the actuators.
    Therefore, all ships are subject to the actuator constraints.
    These input constraints can be divided into magnitude constraint and rate constraint \cite{Doyle1987,Yuan2018}.
    Closed loop systems may become unstable in the case the constraints on the input magnitude are not properly treated \cite{Doyle1987}.
    In the case rate constraint exists, it has been observed that the controlled system may continue to oscillate, which can be understood as a kind of self-excited oscillation, and in the worst case, the system becomes uncontrollable.
    The degradation of control laws due to input constraints is exemplified in \cref{fig:degradation}.
    In the controlled system shown in \cref{fig:degradation}, the same control law was implemented to track the target signal.
    From \cref{fig:degradation}, it can be observed that even if a control law can achieve the tracking of a target signal without input constraints, it may fail in the tracking control in the case it is implemented in a system subject to input constraints.
    This will lead to serious accidents, such as the one illustrated in \cref{fig:miss}.
    Also in the field of aircraft control, oscillation phenomena caused by input rate saturation are known as Pilot-Induced Oscillation (PIO) in Category II \cite{Duda1998} and have been analyzed \cite{Yuan2018}.
    
    Various methods have been studied to control systems with input constraints.
    In the control of a system with input magnitude constraint, the anti-windup technique is well known \cite{Kothare1994,Tarbouriech2009}.
    In the literature \cite{Zhou2006}, a tracking control law was designed for a nonlinear system with an input magnitude constraint and unknown system parameters.
    This study was extended to the system with external disturbance by introducing hyperbolic tangent ($\tanh$) function as a smooth approximation for saturation nonlinearity and using the backstepping method in the literature \cite{Wen2011}.
    In the literature \cite{Wang2018}, stabilizing control laws were designed using SMC for a linear system with constraints on input magnitude and rate.
    In the literature \cite{Sorensen2015}, the constant bearing (CB) guidance \cite{Breivik2006} was applied to bound signals in the controlled system, and it was shown that Multi-Input Multi-Output (MIMO) ship dynamic positioning is possible with smaller input by including CB guidance into the backstepping procedure.
    In the literature \cite{Gaudio2022}, a control law was designed for aircraft maneuvering motion represented by a linear system with elliptical constraints on input magnitude and rate, and it achieved bounded tracking error.
    In the framework of optimal control, constraints on input magnitude and rate can be formulated easily.
    In the literature \cite{Sorensen2017}, in the framework of the dynamic window approach, the constraints of both input magnitude and rate were considered in the computation of feasible velocity states.
    In the literature \cite{Gou2020}, reinforcement learning was utilized to train the path tracking controller for airships.
    In the framework of the reinforcement learning of this work, to treat input constraints, actuator states and their rate were handled as a part of the state variable and the action, respectively.
    Although a variety of efforts can be found as listed here, the authors have not found studies that have designed tracking control laws for nonlinear systems with constrained input magnitude and rate and discussed the convergence of the tracking error.
    
    In the field of ship steering control, some methods to deal with the constraints on rudder manipulation have also been studied.
    In the literature \cite{Witkowska2007}, in addition to the nonlinear maneuvering model, the system of the rudder manipulation was taken into account as a first-order system in the design of the steering control.
    In the literature \cite{Kahveci2013}, adaptive steering control was designed with a linear quadratic controller and Riccati based anti-windup compensator.
    In the literature \cite{Ejaz2017}, SMC was applied to design a steering control for the system with input magnitude constraint, and the asymptotical stability was established.
    Furthermore, in this study, design parameters were adjusted based on fuzzy theory to avoid the chattering of the control signal.
    The work \cite{Du2007} was extended in \cite{Du2017} to the case with the external disturbance and the rudder magnitude constraint.
    In the literature \cite{Zhu2020}, a finite-time adaptive output feedback steering control was designed based on a fuzzy logic system for a nonlinear maneuvering system with input magnitude constraint.
    However, these steering controls did not explicitly address the rate constraint of the rudder manipulation.

    Some reference shaping methods were proposed for the avoidance of input magnitude and rate saturation.
    Reference filter \cite{Fossen2011} makes the reference signal smooth, and, by incorporating saturation elements, explicitly limits the velocity/acceleration of the reference signal.
    In ship course control, for instance, the reference filter makes it possible to avoid actuator magnitude and rate saturation by shaping the reference signal that changes smoothly from the current heading angle to the target heading angle.
    If the reference can be smoothed sufficiently, the performance of the applied control law, e.g., exponential stability, will not be degraded.
    However, the reference signal smoothed by the reference filter does not guarantee that the output of the control law will satisfy the constraints at any state.
    In addition, since the velocity and acceleration are clipped to fixed values, it is not always possible to control the actuator to its limits, in terms of magnitude and rate, considering the current state.
    Therefore, the control method using the reference filter does not allow the actuators to be manipulated to their full extent.
    Reference governor \cite{Bemporad1998} was designed for the controlled system with state and input constraints, and can shape the reference signal for the avoidance of violence of these constraints.
    In this method, nonlinear optimization problems must be solved online, taking into account state constraints in addition to input constraints, and, generally, implementation can be burdensome.
    
    This study focuses on a tracking control law that does not require shaping of the reference signal and guarantees satisfaction of the input magnitude and rate constraints.
    In this study, the authors propose the design of a steering control for a nonlinear ship maneuvering model subject to input magnitude and rate constraints.
    In our method, the tracking problem of the target heading angle with input constraints is transformed into the regulation problem for an error system which is described in a strict-feedback form without any input constraints.
    To derive such a system, the authors introduce hyperbolic tangent ($\tanh$) function and auxiliary variables to deal with the constraints on input as some existing method \cite{Wen2011,Wang2013,Zheng2018,Zhu2020}.
    Furthermore, by a time derivative of the formulated variable, due to the feature of the derivative of $\tanh$ function, an auxiliary system for rudder manipulation is constructed in the strict-feedback form.
    Using this technique two times, for the constraints of magnitude and rate respectively, both actuator constraints are successfully incorporated into the cascade system which does not have any input constraints.
    The steering control is designed using the backstepping technique \cite{Krstic1995,Fossen1999BC} for the resulting strict-feedback system.
    In our method, it is shown that, for the feasible target signals, the tracking error exponentially converges to zero.
    Although the proposed steering control has a limitation in terms of numerical implementation, it is the first attempt at the tracking control for nonlinear systems under input magnitude and rate saturations with exponential convergence.
    To verify the proposed control law, numerical experiments are conducted.
    
    The rest of the paper is organized as follows:
    \cref{sec:notation} describes the notation used in this manuscript;
    \cref{sec:problem} describes the tracking problem of the target heading angle considered in this study;
    \cref{sec:design} describes the conventional method and the design procedure of the proposed control law;
    \cref{sec:experiment} describes the numerical experiments implemented to verify the proposed steering control and compare the performance with the conventional method;
    \cref{sec:discussion} discusses the property of the proposed method in terms of the unboundedness of control signal point of view;
    finally, \cref{sec:conclusion} concludes the study.

\section{Notation}
    \label{sec:notation}
    $\realset$ represents the set of all real numbers.
    $\mathbb{R}^n$ represents the $n$-dimensional Euclidean space.
    $\realset_{+}$ represents the set of all positive real numbers.
    $|x|$ represents the absolute value of $x \in \mathbb{R}$.
    The overdot `` $\dot{}$ '' represents the derivative with respect to time $t$.
    The saturation function $\sat(s, \bar{s}) : \realset \times \realset_{+} \rightarrow [-\bar{s}, \bar{s}]$ is defined as:
    \begin{equation}
        \sat(s, \bar{s})
            :=
                \left\{
                    \begin{alignedat}{2}
                        &s \quad  && \text{for} \quad | s | \leq \bar{s}  \enspace ,  \\
                        &\sgn(s) \bar{s} \quad && \text{for} \quad |s| > \bar{s}  \enspace .
                    \end{alignedat}
                \right .
    \end{equation}
    $\diag(a_{1}, \cdots, a_{n})$ represents a diagonal matrix $A \in \realset^{n \times n}$ such that:
    \begin{equation}
        A_{ij} :=
                \left\{
                    \begin{alignedat}{2}
                        &a_{i} \qquad &&\text{for} ~ i = j  \enspace ,  \\
                        &0 \qquad &&\text{for} ~ i \neq j  \enspace .
                    \end{alignedat}
                \right .
    \end{equation}

\section{Problem formulation}
    \label{sec:problem}

    \subsection{Maneuvering model}

\begin{figure}[tb]
            \centering
            \includegraphics[width=0.4\hsize]{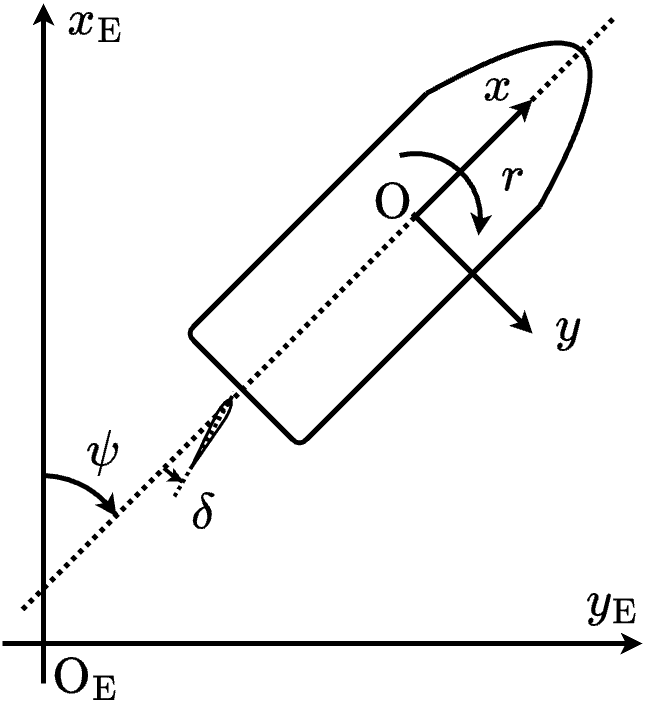}
            \caption{Coordinate systems.}
            \label{fig:coordinate}
        \end{figure}

        The ship is assumed to move on an Earth-fixed coordinate $\mathrm{O}_{\mathrm{E}} - x_{\mathrm{E}} y_{\mathrm{E}}$ as \cref{fig:coordinate} shows.
        $\psi(t)$, $r(t)$, and $\delta(t)$ represent the heading angle, yawing angular velocity, and rudder angle, respectively.
        $\mathrm{O}-xy$ in \cref{fig:coordinate} represents a body-fixed coordinate system with the origin on the center of the ship.
    
        The heading angle $\psi(t)$ and the yawing angular velocity $r(t)$ follow \cref{eq:kinematics}.
        \begin{equation}
            \label{eq:kinematics}
            \dot{\psi}(t) = r(t)\enspace .
        \end{equation}
        In this study, it is assumed that the change in ship speed due to turning motion is insignificant.
        Thus the ship speed is assumed to be constant.
        In this situation, it is known that the maneuvering motion of a ship can be modeled by the following single-input single-output (SISO) state equation:
        \begin{equation}
            \label{eq:dynamics}
            \dot{r}(t) = f(r(t)) + b \delta(t)
            \enspace ,
        \end{equation}
        where $f : \realset \rightarrow \realset$ is a two times differentiable function, $b \neq 0$.
        Well-known examples of this formulation are the linear system (Nomoto's KT model) \cite{Nomoto1957}:
        \begin{equation}
            \begin{aligned}
                & T \dot{r}(t) + r(t) = K \delta(t)  \\
                & \Leftrightarrow \dot{r}(t) = -\frac{1}{T} r(t) + \frac{K}{T} \delta(t)  \enspace ,
            \end{aligned}
        \end{equation}
        and the system expressed by three-dimensional polynomial of $r(t)$ \cite{Norrbin1963}:
        \begin{equation}
            \label{eq:Norrbin}
            \begin{aligned}
                & T \dot{r}(t) + H(r) = K \delta(t)  \\
                & \Leftrightarrow \dot{r}(t) = -\frac{1}{T} H(r(t)) + \frac{K}{T} \delta(t) \enspace ,
            \end{aligned}
        \end{equation}
        where $K \neq 0, ~ T \neq 0$, $H : \realset \rightarrow \realset$ is a function of $r$ which is defined as:
        \begin{equation}
            H(r) = n_{3} r^3 + n_{2} r^2 + n_{1} r + n_{0}
        \end{equation}
        with constants $n_{i} \in \realset ~ (i = 0, 1, 2, 3)$.

    \subsection{Constraints on rudder manipulation}

        It is customary for actual control systems, including ships, to have constraints on their input.
        In this study, as constraints on rudder angle $\delta(t)$ and rudder manipulation speed $\dot{\delta}(t)$, the following inequalities are considered:
        \begin{equation}
            \label{eq:magnitude_saturation}
            | \delta(t) | \leq M
            \quad \forall t
            \enspace ,
        \end{equation}
        \begin{equation}
            \label{eq:rate_saturation}
            | \dot{\delta}(t) | \leq R
            \quad \forall t
            \enspace ,
        \end{equation}
        where $M > 0$ and $R > 0$ are constants.

        The formulations \cref{eq:magnitude_saturation,eq:rate_saturation} are reasonable as constraints imposed on the rudder manipulation system of ships.
        In a typical rudder manipulation system, the rudder angle is restricted to an interval symmetrical from the origin, for instance to $[ -35, 35 ]$ degree, which can be expressed by \cref{eq:magnitude_saturation}.
        The constraint on the rudder manipulation speed must be expressed in the formulation that it always does not exceed a certain threshold value. In the design procedure of ship controllers, the constraint on rudder manipulation speed is often treated by introducing a first-order system \cite{Shouji1992,Witkowska2007} of rudder angle $\delta(t)$ with the commanded rudder angle $\delta_{\mathrm{c}}(t)$ as the input:
        \begin{equation}
            \label{eq:1st_order_rudder_dynamics}
            \dot{\delta}(t) = \frac{1}{T_{\mathrm{R}}} ( K_{\mathrm{R}} \delta_{\mathrm{c}}(t) - \delta(t) ) \enspace ,
        \end{equation}
        where $T_{\mathrm{R}} > 0$ and $K_{\mathrm{R}} > 0$ are constants.
        Under this formulation, $\dot{\delta}(t)$ depends on the deviation between the rudder angle $\delta(t)$ and the commanded rudder angle $\delta_{\mathrm{c}}(t)$.
        Therefore the values of $T_{\mathrm{R}}$ and $K_{\mathrm{R}}$ must be adjusted to moderate the rudder manipulation speed to guarantee the satisfaction of constraint \cref{eq:rate_saturation}.
        However, if the rudder manipulation is slowed to the extent that the satisfaction of constraint \cref{eq:rate_saturation} is guaranteed for any $ | \delta(t) | \leq M $ and $| \delta_{\mathrm{c}}(t) | \leq M$, the response of $\delta(t)$ will be too slow that the model is inappropriate.
        In the proposed method, the constraint \cref{eq:rate_saturation} is directly addressed instead of assuming the first-order system of rudder manipulation \cref{eq:1st_order_rudder_dynamics}.

    \subsection{Desired heading angle}

        The target heading angle $\psid(t)$ is assumed to be given as the function of time $t$. Here it is assumed that $\psid(t)$ is four times differentiable.

        It is assumed that the time series $\psid(t)$ is feasible under constraints \cref{eq:kinematics,eq:dynamics,eq:magnitude_saturation,eq:rate_saturation}.
        For instance, if $\psid(t)$ includes an oscillation with high frequency, the exponential stabilization of the tracking error for this $\psid(t)$ is unachievable.
        Thus, such a $\psid(t)$ is out of the scope of this study.
        The condition for $\psid(t)$ to be feasible is derived as follows.
        \Cref{eq:dynamics} gives:
        \begin{equation}
            \delta(t) = \frac{\dot{r}(t) - f(r(t))}{b}
            \enspace ,
        \end{equation}
        \begin{equation}
            \dot{\delta}(t)
                = \frac{1}{b} \Big(
                    \ddot{r}(t) - \frac{\mathrm{d} f}{\mathrm{d} r}(r(t)) \dot{r}(t)
                \Big)
            \enspace .
        \end{equation}
        With these, canceling $\delta(t)$ and $\dot{\delta}(t)$ in \cref{eq:magnitude_saturation,eq:rate_saturation}, the followings are obtained:
        \begin{equation}
            \label{eq:m_sat_r}
            \left|
                \frac{\dot{r}(t) - f(r(t))}{b}
            \right|
            \leq M
            \quad \forall t
            \enspace ,
        \end{equation}
        \begin{equation}
            \label{eq:r_sat_r}
            \left|
                \frac{1}{b} \big(
                    \ddot{r}(t) - \frac{\mathrm{d} f}{\mathrm{d} r}(r(t)) \dot{r}(t)
                \big)
            \right|
            \leq R
            \quad \forall t
            \enspace ,
        \end{equation}
        Now $\rd(t) := \dot{\psi}^{\mathrm{d}}(t)$ is defined.
        In \cref{eq:m_sat_r,eq:r_sat_r}, letting $r(t) \leftarrow \rd(t)$, $\dot{r}(t) \leftarrow \dot{r}^{\mathrm{d}}(t)$, $\ddot{r}(t) \leftarrow \ddot{r}^{\mathrm{d}}(t)$, the conditions on $\rd(t)$ are obtained as:
        \begin{equation}
            \label{eq:m_sat_rd}
            \left|
                \frac{\dot{r}^{\mathrm{d}}(t) - f(\rd(t))}{b}
            \right|
            \leq M
            \quad \forall t
            \enspace ,
        \end{equation}
        \begin{equation}
            \label{eq:r_sat_rd}
            \left|
                \frac{1}{b} \big(
                    \ddot{r}^{\mathrm{d}}(t) - \frac{\mathrm{d} f}{\mathrm{d} r} (\rd(t)) \dot{r}^{\mathrm{d}}(t)
                \big)
            \right|
            \leq R
            \quad \forall t
            \enspace .
        \end{equation}
        \Cref{eq:m_sat_rd,eq:r_sat_rd} are the necessary conditions for the exact tracking of $\psid(t)$.
        Thus $\psid(t)$ that does not satisfy \cref{eq:m_sat_rd,eq:r_sat_rd} is out of the scope of this study.

    \subsection{Control objective}

        The tracking error:
        \begin{equation}
            \epsi(t) := \psi(t) -\psid(t)
        \end{equation}
        is defined.
        The goal of the control law designed in this study is to make the tracking error $\epsi(t)$ exponentially stable \cite{Khalil1992} at the origin.

\section{Control design}
\label{sec:design}

    In this section, the design procedure of the proposed steering control is described.
    The authors first describe a conventional method and its problem in \cref{sec:conventional}.
    This is designed by considering a cascade system composed of kinematics \cref{eq:kinematics}, dynamics \cref{eq:dynamics}, and sometimes a rudder manipulation system \cref{eq:1st_order_rudder_dynamics}.
    Next, in \cref{sec:system}, the authors propose the expression of state variables with $\tanh$ function and auxiliary variables to guarantee the satisfaction of input constraints.
    Moreover, based on this expression, an unconstrained strict-feedback system \cite{Krstic1995} is derived.
    Then, in \cref{sec:autopilot}, the proposed steering control for $\psid(t)$ satisfying \cref{eq:m_sat_rd,eq:r_sat_rd} is constructed based on the backstepping method \cite{Krstic1995,Fossen1999BC}, and the exponential stability is proven in \cref{sec:stability}.
    
    It should be noted that, in the proposed method, it is assumed that the tracking of $\psid(t)$ is possible with mild rudder manipulation.
    This point is detailed in \cref{sec:discussion}.

    In the following, $(t)$, which indicates the dependence of variables on time, is omitted to simplify the description.

    \subsection{Conventional method}
    \label{sec:conventional}

        Without input constraints \cref{eq:magnitude_saturation,eq:rate_saturation}, it is known that a steering control that achieves the control objective, i.e., exponentially stabilizes the tracking error at the origin, can be designed by applying the backstepping method to cascade systems, for instance, kinematics \cref{eq:kinematics} and dynamics \cref{eq:dynamics}.
        This method is described below.
        The time derivative of $\epsi$ is calculated with \cref{eq:kinematics} as:
        \begin{equation}
            \label{eq:epsi_dot}
            \dot{e}_{\psi} = r - \dot{\psi}^{\mathrm{d}}
            \enspace .
        \end{equation}
        Here an error variable $\er := r - \{ -c_{1} \epsi - (-\dot{\psi}^{\mathrm{d}} ) \}$ is defined with $c_{1} > 0$. With $\er$, \cref{eq:epsi_dot} is written as:
        \begin{equation}
            \label{eq:epsi_dot_2}
            \dot{e}_{\psi} = -c_{1} \epsi + \er
            \enspace .
        \end{equation}
        The time derivative of $\er$ is calculated as:
        \begin{equation}
            \label{eq:er_dot}
            \dot{e}_{r} = f(r) + b \delta + c_{1} (r - \dot{\psi}^{\mathrm{d}}) - \ddot{\psi}^{\mathrm{d}}
            \enspace .
        \end{equation}
        The control law:
        \begin{equation}
            \delta = \alpha_{\delta}( \psi, r, \psid, \dot{\psi}^{\mathrm{d}}, \ddot{\psi}^{\mathrm{d}} )
        \end{equation}
        is designed as:
        \begin{equation}
            \label{eq:simple_bs}
            \begin{split}
                & \alpha_{\delta}( \psi, r, \psid, \dot{\psi}^{\mathrm{d}}, \ddot{\psi}^{\mathrm{d}} )  \\
                    &\quad := \frac{1}{b}
                        \big[
                            -c_{2} \er - \epsi
                                - \big\{
                                    f(r) + c_{1} (r - \ddot{\psi}^{\mathrm{d}} ) \ddot{\psi}^{\mathrm{d}}
                                \big\}
                        \big]
            \end{split}
        \end{equation}
        with $c_{2} > 0$.
        Then \cref{eq:er_dot} becomes:
        \begin{equation}
            \label{eq:er_dot_2}
            \dot{e}_{r} = -c_{2} \er - \epsi
            \enspace .
        \end{equation}

        Now an error variable:
        \begin{equation}
            e := ( \epsi ~~ \er )^{\top} \in \realset^{2}
        \end{equation}
        is defined.
        In addition, a function $V_{e} : \realset^{2} \rightarrow \realset$ is defined as:
        \begin{equation}
            \label{eq:V_e}
            V_{e} := \frac{1}{2} e^{\top} e
            > 0, \quad \forall e \neq 0
            \enspace .
        \end{equation}
        With \Cref{eq:epsi_dot_2,eq:er_dot_2}, the system for $e$ is derived as:
        \begin{equation}
            \label{eq:e_dot}
            \dot{e} = -C_{e} e + S_{e} e
            \enspace ,
        \end{equation}
        where
        \begin{equation}
            C_{e} := \diag(c_{1}, c_{2})
            \enspace ,
        \end{equation}
        \begin{equation}
            S_{e} :=
                \begin{pmatrix}
                    0 & 1  \\
                    -1 & 0  \\
                \end{pmatrix}
            \enspace .
        \end{equation}
        With \cref{eq:e_dot}, the time derivative of $V_{e}$ is calculated as:
        \begin{equation}
            \label{eq:V_e_dot}
            \begin{split}
                \dot{V}
                    &= e^{\top} \dot{e}  \\
                    &= e^{\top} ( -C_{e} e + S_{e} e )  \\
                    &= -e^{\top} C_{e} e < 0 \quad \forall e \neq 0
                \enspace .
            \end{split}
        \end{equation}
        Now it is shown that $V_{e}$ is a global Lyapunov function \cite{Khalil1992} on $\realset^{2}$.
        That is, if no constraints are imposed on the input, this method can make $e$ globally exponentially stable at the origin.
        Thus, the goal of the control design has been achieved.

        However, the steering control $\alphadelta$ may not achieve the control objective if it is implemented in the system with constraints \cref{eq:magnitude_saturation,eq:rate_saturation}.
        $\alphadelta$ may output the command that does not satisfy the \cref{eq:magnitude_saturation,eq:rate_saturation}.
        In many cases, for a given command from $\alphadelta$, $\delta$ and $\dot{\delta}$ are determined based on:
        \begin{equation}
            \label{eq:sat_function_m}
            \delta = \sat(\alphadelta(\cdot), M)
            \enspace ,
        \end{equation}
        \begin{equation}
            \label{eq:sat_function_r}
            \dot{\delta} = \sat(\dot{\alpha}_{\delta}(\cdot), R)
            \enspace .
        \end{equation}
        In the case saturation occurs in these processes, the desired performance may not be achieved as indicated in \cite{Doyle1987,Yuan2018}.

    \subsection{Auxiliary system for input constraints}
    \label{sec:system}

        To deal with \cref{eq:magnitude_saturation}, $\tanh$ function and an auxiliary variable are introduced, as some existing method \cite{Wen2011,Wang2013,Zheng2018,Zhu2020}, and an auxiliary system is derived.
        The rudder angle is expressed as:
        \begin{equation}
            \label{eq:u_tanh}
            \delta = M \tanh (\kdelta \deltati )
            \enspace ,
        \end{equation}
        with $\kdelta > 0$ and an auxiliary variable $\deltati \in \realset$.
        The time derivative of \cref{eq:u_tanh} is calculated, using the fundamental feature of $\tanh$ function, as:
        \begin{equation}
            \label{eq:u_dot_1}
            \begin{aligned}
                \dot{\delta}
                    &= M ( 1 - \tanh^{2}(\kdelta \deltati ) ) \kdelta \dot{\deltati}  \\
                    &= \kdelta M \left\{
                        1 - \left( \frac{ \delta }{M} \right)^{2}
                    \right\} \dot{\deltati}  \\
                    &= \kdelta \frac{M^{2} - \delta^{2} }{M} \dot{\deltati} \enspace .
            \end{aligned}
        \end{equation}
        Defining a function $g_{\delta} : \realset \rightarrow \realset$ as:
        \begin{equation}
            g_{\delta}(\delta) := \kdelta \frac{M^{2} - \delta^{2}}{M}
            \enspace ,
        \end{equation}
        \cref{eq:u_dot_1} becomes:
        \begin{equation}
            \label{eq:u_dot_2}
            \dot{\delta} = g_{\delta}(\delta ) \dot{\deltati}
            \enspace .
        \end{equation}
        Here new auxiliary state variable $\xi := \dot{\deltati}$ is introduced.
        In the following, it is assumed that the value $M^{2} - \delta^{2} > 0$ is enough large, that is the value of $g_{\delta}(\delta)$ is enough larger than zero.
        This assumption is for the avoidance of the numerical overflow in the controlled system, which is detailed in \cref{sec:discussion}.

        With \cref{eq:u_dot_2}, the constraint on rudder manipulation speed \cref{eq:rate_saturation} is converted as:
        \begin{equation}
            \label{eq:saturation_xi}
            \begin{aligned}
                & \kdelta \frac{M^{2} - \delta^{2} }{M} | \xi | \leq R \quad \forall t  \\
                \Leftrightarrow
                \quad
                & | \xi | \leq \frac{M R}{\kdelta (M^{2} - \delta^{2} ) } \quad \forall t \enspace .
            \end{aligned}
        \end{equation}
        To guarantee the satisfaction of the constraint on $\xi$ \cref{eq:saturation_xi}, $\tanh$ function and an auxiliary variable are again introduced. $\xi$ is expressed as:
        \begin{equation}
            \label{eq:xi_tanh}
            \xi = \frac{M R}{\kdelta (M^{2} - \delta^{2} ) } \tanh( \kxi \xiti )
            \enspace ,
        \end{equation}
        with $\kxi > 0$ and an auxiliary variable $\xiti \in \realset$.
        The time derivative of \cref{eq:xi_tanh} is calculated as:
        \begin{equation}
            \label{eq:xi_dot}
            \begin{aligned}
                \dot{\xi}
                    &=
                        \frac
                            {\mathrm{d}}
                            {\mathrm{d} \delta}
                                \bigg\{
                                    \frac{M R}{\kdelta (M^{2} - \delta^{2} ) }
                                \bigg\}
                        g_{\delta}( \delta ) \xi
                        \times \frac{\kdelta (M^{2} - \delta^{2} ) }{M R} \xi  \\
                        & \quad + \frac{M R}{\kdelta (M^{2} - \delta^{2} ) }
                        \big( 1 - \tanh^{2}(\kxi \xiti ) \big) \kxi \dot{\xiti}  \\
                    &=
                        \frac{2 \kdelta \delta \xi^{2} }{M}  \\
                        & \quad
                            + \frac{ \kdelta \kxi (M^{2} - \delta^{2} ) }{M R}
                            \left[
                                \bigg\{
                                    \frac{M R}{ \kdelta ( M^{2} - \delta^{2} ) }
                                \bigg\}^{2}
                                - \xi^{2}
                            \right]
                            \dot{\xiti}  \\
                    &=
                        f_{\xi}( \delta, \xi )
                        + g_{\xi}( \delta, \xi ) \dot{\xiti} \enspace ,
            \end{aligned}
        \end{equation}
        where
        \begin{equation}
            \left\{
                \begin{aligned}
                    f_{\xi}( \delta, \xi )
                        &:= \frac{2 \kdelta \delta \xi^{2} }{M}  \\
                    g_{\xi}( \delta, \xi )
                        &:= \frac{ \kdelta \kxi (M^{2} - \delta^{2} ) }{M R}
                            \left[
                                \bigg\{
                                    \frac{M R}{ \kdelta ( M^{2} - \delta^{2} ) }
                                \bigg\}^{2}
                                - \xi^{2}
                            \right]
                \end{aligned}
            \right.\enspace .
        \end{equation}
        Here new auxiliary variable $\eta := \dot{\xiti}$ is introduced as the input.
        In the following, it is assumed that the value $ [ M R / \{ \kdelta ( M^{2} - \delta^{2} ) \} ]^{2} - \xi^{2}$ is enough large, that is, the value of $g_{\xi}( \delta, \xi )$ is enough larger than zero.
        This assumption is also for the avoidance of the numerical overflow in the controlled system, which is detailed in \cref{sec:discussion}.

        Now the whole system with the states $\psi$, $r$, $\delta$, $\xi$ and the input $\eta$ is described as a cascade system:
        \begin{equation}
            \label{eq:state_eqs}
            \left \{
                \begin{aligned}
                    \dot{\psi} &= r  \\
                    \dot{r} &= f(r) + b \delta  \\
                    \dot{\delta} &= g_{\delta}(\delta) \xi  \\
                    \dot{\xi} &=
                        f_{\xi}(\delta, \xi) + g_{\xi}(\delta, \xi) \eta  \enspace .
                \end{aligned}
            \right .
        \end{equation}
        It should be noted that:
        \begin{equation}
            \left \{
                \begin{aligned}
                    g_{\delta}(\delta)  &\neq 0  \\
                    g_{\xi}(\delta, \xi) &\neq 0  \\
                \end{aligned}
            \right .
            \quad \forall t
            \enspace .
        \end{equation}
        This is because we have
        \begin{equation}
            | \delta | < M
            \quad \forall t
            \enspace ,
        \end{equation}
        \begin{equation}
            | \dot{\delta} | < R
            \quad \forall t
            \enspace ,
        \end{equation}
        due to \cref{eq:u_tanh,eq:xi_tanh}, respectively.
        This means that the satisfaction of constraints \cref{eq:magnitude_saturation,eq:rate_saturation} is guaranteed in \cref{eq:state_eqs}.
        In addition, \cref{eq:state_eqs} has the strict-feedback form \cite{Krstic1995}, where all the state equations have the input-affine form and are described by state variables that appear above and input.
        Here the problem defined in \cref{sec:problem} is transformed into the tracking problem for \cref{eq:state_eqs} without any constraints on input $\eta$.

    \subsection{Design of steering control}
    \label{sec:autopilot}

        In this section, the design of the proposed steering control $\eta = \alphaeta(\cdot)$ is described.
        Due to the feature of the introduced cascade system \cref{eq:state_eqs}, $\alphaeta$ can be designed using the backstepping method \cite{Krstic1995,Fossen1999BC}.
        In the following, the arguments of functions: $r$, $\delta$, and $\xi$ are omitted to simplify the description.

        The first error variable is chosen as:
        \begin{equation}
            \label{eq:z1_def}
            z_{1} := \epsi
            \enspace .
        \end{equation}
        The time derivative of \cref{eq:z1_def} yields:
        \begin{equation}
            \label{eq:dot_z1_1}
            \dot{z}_{1} = r - \dot{\psi}^{\mathrm{d}}
            \enspace .
        \end{equation}
        Here a new error variable is defined as:
        \begin{equation}
            \label{eq:z2_def}
            \begin{split}
                z_{2}
                    & := r - \big\{ -c_{1} z_{1} - (-\dot{\psi}^{\mathrm{d}}) \big\}  \\
                    & = c_{1} ( \psi - \psid ) + r - \dot{\psi}^{\mathrm{d}} \enspace ,
            \end{split}
        \end{equation}
        with a design parameter $c_{1} > 0$.
        Using $z_{2}$, \cref{eq:dot_z1_1} becomes:
        \begin{equation}
            \label{eq:dot_z1_2}
            \dot{z}_{1} = -c_{1} z_{1} + z_{2}
            \enspace .
        \end{equation}
        The time derivative of \cref{eq:z2_def} yields:
        \begin{equation}
            \label{eq:dot_z2_1}
            \dot{z}_{2}
                = c_{1} ( r - \dot{\psi}^{\mathrm{d}} ) + f + b \delta - \ddot{\psi}^{\mathrm{d}}
            \enspace .
        \end{equation}
        Here a new error variable is defined as:
        \begin{equation}
            \label{eq:z3_def}
            \begin{split}
                z_{3}
                    & :=
                        b \delta
                        - \Big[
                            -c_{2} z_{2} - z_{1}
                            - \big\{
                                c_{1} ( r - \dot{\psi}^{\mathrm{d}} ) + f - \ddot{\psi}^{\mathrm{d}}
                            \big\}
                        \Big]  \\
                    & = ( c_{1} c_{2} + 1 ) ( \psi - \psid ) + ( c_{1} + c_{2} ) ( r - \dot{\psi}^{\mathrm{d}} )  \\
                        & \qquad + f + b \delta - \ddot{\psi}^{\mathrm{d}} \enspace ,
            \end{split}
        \end{equation}
        with a design parameter $c_{2} > 0$.
        Using $z_{3}$, \cref{eq:dot_z2_1} becomes:
        \begin{equation}
            \label{eq:dot_z2_2}
            \dot{z}_{2} = -c_{2} z_{2} - z_{1} + z_{3}
            \enspace .
        \end{equation}
        The time derivative of \cref{eq:z3_def} yields:
        \begin{equation}
            \label{eq:dot_z3_1}
            \begin{split}
                \dot{z}_{3}
                    & = ( c_{1} c_{2} + 1 ) ( r - \dot{\psi}^{\mathrm{d}} )
                        + ( c_{1}+c_{2} ) \big(
                            f + b \delta - \ddot{\psi}^{\mathrm{d}}
                        \big)  \\
                    & \quad + \frac{\mathrm{d} f}{\mathrm{d} r} ( f + b \delta ) + b g_{\delta} \xi - \dddot{\psi}^{\mathrm{d}} \enspace .
            \end{split}
        \end{equation}
        Here a new error variable is defined as:
        \begin{equation}
            \label{eq:z4_def}
            \begin{split}
                z_{4}
                    & :=
                        b g_{\delta} \xi
                        - \bigg[
                            -c_{3} z_{3} - z_{2}  \\
                            & \qquad - \Big\{
                                ( c_{1} c_{2} + 1 ) ( r - \dot{\psi}^{\mathrm{d}} )  \\
                                & \quad \quad \quad + ( c_{1}+c_{2} ) \big(
                                    f + b \delta - \ddot{\psi}^{\mathrm{d}}
                                \big)  \\
                                & \quad \quad \quad + \frac{\mathrm{d} f}{\mathrm{d} r} ( f + b \delta ) - \dddot{\psi}^{\mathrm{d}}
                            \Big\}
                        \bigg]  \\
                    & =
                        ( c_{1} + c_{3} + c_{1} c_{2} c_{3} ) ( \psi - \psid )  \\
                        & \qquad + ( c_{1} c_{2} + c_{2} c_{3} + c_{3} c_{1} + 2 ) ( r - \dot{\psi}^{\mathrm{d}} )  \\
                        & \qquad + ( c_{1} + c_{2} + c_{3} ) ( f + b \delta - \ddot{\psi}^{\mathrm{d}} )  \\
                        & \qquad + \frac{\mathrm{d} f}{\mathrm{d} r} ( f + b \delta ) + b g_{\delta} \xi - \dddot{\psi}^{\mathrm{d}} \enspace ,
            \end{split}
        \end{equation}
        with a design parameter $c_{3} > 0$.
        Using $z_{4}$, \cref{eq:dot_z3_1} becomes:
        \begin{equation}
            \label{eq:dot_z3_2}
            \dot{z}_{3} = -c_{3} z_{3} - z_{2} + z_{4}
            \enspace .
        \end{equation}
        The time derivative of \cref{eq:z4_def} yields:
        \begin{equation}
            \label{eq:dot_z4_1}
            \begin{split}
                \dot{z}_{4}
                    & = ( c_{1} + c_{3} + c_{1} c_{2} c_{3} ) ( r - \dot{\psi}^{\mathrm{d}} )  \\
                    & \quad + ( c_{1} c_{2} + c_{2} c_{3} + c_{3} c_{1} + 2 ) ( f + b \delta - \ddot{\psi}^{\mathrm{d}} )  \\
                    & \quad + ( c_{1} + c_{2} + c_{3} )
                        \Big\{
                            \frac{\mathrm{d} f}{\mathrm{d} r} ( f + b \delta ) + b g_{\delta} \xi
                            - \dddot{\psi}^{\mathrm{d}}
                        \Big\}  \\
                    & \quad + \frac{\mathrm{d}^{2} f}{\mathrm{d} r^{2}} ( f + b \delta )^{2}
                        + \frac{\mathrm{d} f}{\mathrm{d} r}
                            \Big\{
                                \frac{\mathrm{d} f}{\mathrm{d} r} ( f + b \delta ) + b g_{\delta} \xi
                            \Big\}  \\
                    & \quad + b
                        \Big\{
                            \frac{\mathrm{d} \gdelta}{\mathrm{d} \delta} \gdelta \xi^{2}
                            + \gdelta ( \fxi + \gxi \eta )
                        \Big\}
                        - \ddddot{\psi}^{\mathrm{d}} \enspace .
            \end{split}
        \end{equation}
        Here the steering control is designed as:
        \begin{equation}
            \label{eq:alpha_eta}
            \begin{split}
                \alphaeta
                    & =
                        \frac{1}{b g_{\delta} g_{\xi}}
                            \bigg( -c_{4} z_{4} - z_{3}  \\
                                & \quad - \bigg [
                                    ( c_{1} + c_{3} + c_{1} c_{2} c_{3} ) ( r - \dot{\psi}^{\mathrm{d}} )  \\
                                    & \quad \quad + ( c_{1} c_{2} + c_{2} c_{3} + c_{3} c_{1} + 2 ) ( f + b \delta - \ddot{\psi}^{\mathrm{d}} )  \\
                                    & \quad \quad + ( c_{1} + c_{2} + c_{3} )
                                        \Big\{
                                            \frac{\mathrm{d} f}{\mathrm{d} r} ( f + b \delta ) + b g_{\delta} \xi
                                            - \dddot{\psi}^{\mathrm{d}}
                                        \Big\}  \\
                                    & \quad \quad + \frac{\mathrm{d}^{2} f}{\mathrm{d} r^{2}} ( f + b \delta )^{2}
                                        + \frac{\mathrm{d} f}{\mathrm{d} r}
                                            \Big\{
                                                \frac{\mathrm{d} f}{\mathrm{d} r} ( f + b \delta ) + b g_{\delta} \xi
                                            \Big\}  \\
                                    & \quad \quad + b
                                        \Big\{
                                            \frac{\mathrm{d} \gdelta}{\mathrm{d} \delta} \gdelta \xi^{2}
                                            + \gdelta \fxi
                                        \Big\}
                                        - \ddddot{\psi}^{\mathrm{d}}
                                \bigg ]
                            \bigg)  \\
                    & =
                        \frac{1}{b g_{\delta} g_{\xi}}
                            \bigg(  \\
                                & \qquad -( c_{1} c_{2} + c_{3} c_{4} + c_{4} c_{1}  \\
                                    & \qquad \qquad + c_{1} c_{2} c_{3} c_{4} + 1 ) ( \psi - \psid )  \\
                                & \qquad - ( 2 c_{1} + c_{2} + c_{3} + 2 c_{4} + c_{1} c_{2} c_{3}  \\
                                    & \qquad \qquad + c_{4} c_{1} c_{2} + c_{3} c_{4} c_{1} + c_{2} c_{3} c_{4} ) ( r - \dot{\psi}^{\mathrm{d}} )  \\
                                & \qquad - ( c_{1} c_{2} + c_{1} c_{3} + c_{1} c_{4}  \\
                                    & \qquad \qquad + c_{2} c_{3} + c_{2} c_{4} + c_{3} c_{4} + 3 ) ( f + b \delta - \ddot{\psi}^{\mathrm{d}} )  \\
                                & \qquad - ( c_{1} + c_{2} + c_{3} + c_{4} )  \\
                                    & \qquad \quad
                                        \times \Big\{
                                            \frac{\mathrm{d} f}{\mathrm{d} r} ( f + b \delta ) + b g_{\delta} \xi - \dddot{\psi}^{\mathrm{d}}
                                        \Big\}  \\
                                & \qquad
                                    - \Big[
                                        ~ \frac{\mathrm{d}^{2} f}{\mathrm{d} r^{2}} ( f + b \delta )^{2}
                                        + \frac{\mathrm{d} f}{\mathrm{d} r}
                                            \Big\{
                                                \frac{\mathrm{d} f}{\mathrm{d} r} ( f + b \delta ) + b g_{\delta} \xi
                                            \Big\}  \\
                                        & \qquad \quad+ b
                                        \Big\{
                                            \frac{\mathrm{d} \gdelta}{\mathrm{d} \delta} \gdelta \xi^{2}
                                            + \gdelta \fxi
                                        \Big\}
                                        - \ddddot{\psi}^{\mathrm{d}}
                                    \Big]
                            \bigg) \enspace ,
            \end{split}
        \end{equation}
        with a design parameter $c_{4} > 0$.
        Substituting \cref{eq:alpha_eta} into \cref{eq:dot_z4_1}, it becomes:
        \begin{equation}
            \label{eq:dot_z4_2}
            \dot{z}_{4} = -c_{4} z_{4} - z_{3}
            \enspace .
        \end{equation}

    \subsection{Exponential stability}
    \label{sec:stability}

        In this section, the exponential stability of the tracking error at the origin ($\epsi = 0$) is presented for the feasible target signal.

        The error variable $z := ( z_{1} \; z_{2} \; z_{3} \; z_{4} )^{\top}$ and a candidate Lyapunov function:
        \begin{equation}
            \label{eq:V}
            V( z ) := \frac{1}{2} z^{\top} z
            > 0 \quad \forall z \neq 0
        \end{equation}
        are defined.
        With \cref{eq:dot_z1_2,eq:dot_z2_2,eq:dot_z3_2,eq:dot_z4_2}, the system of $z$ is described as:
        \begin{equation}
            \label{eq:dot_z}
            \dot{z} = -C z + S z
            \enspace ,
        \end{equation}
        where
        \begin{equation}
            C := \diag(c_{1}, c_{2}, c_{3}, c_{4})
            \enspace ,
        \end{equation}
        \begin{equation}
            S :=
                \begin{pmatrix}
                    0 & 1 & 0 & 0  \\
                    -1 & 0 & 1 & 0  \\
                    0 & -1 & 0 & 1  \\
                    0 & 0 & -1 & 0
                \end{pmatrix}
            \enspace .
        \end{equation}
        Therefore the time derivative of $V$ is:
        \begin{equation}
            \label{eq:dot_V}
            \begin{split}
                \dot{V}
                    &= z^{\top} \dot{z}  \\
                    &= z^{\top} (-C z + S z)  \\
                    &= -z^{\top} C z < 0 \quad \forall z \neq 0
                \enspace .
            \end{split}
        \end{equation}
        Thus, it is proven that, if \cref{eq:m_sat_rd,eq:r_sat_rd} are satisfied, then $z$ is locally uniformly exponentially stable at the origin.

\section{Numerical experiments}
\label{sec:experiment}

    The proposed method was verified in the numerical experiments of the target heading angle tracking control.

    \subsection{Setting}

\begin{figure}[tb]
            \centering
            \includegraphics[width=1.0\hsize]{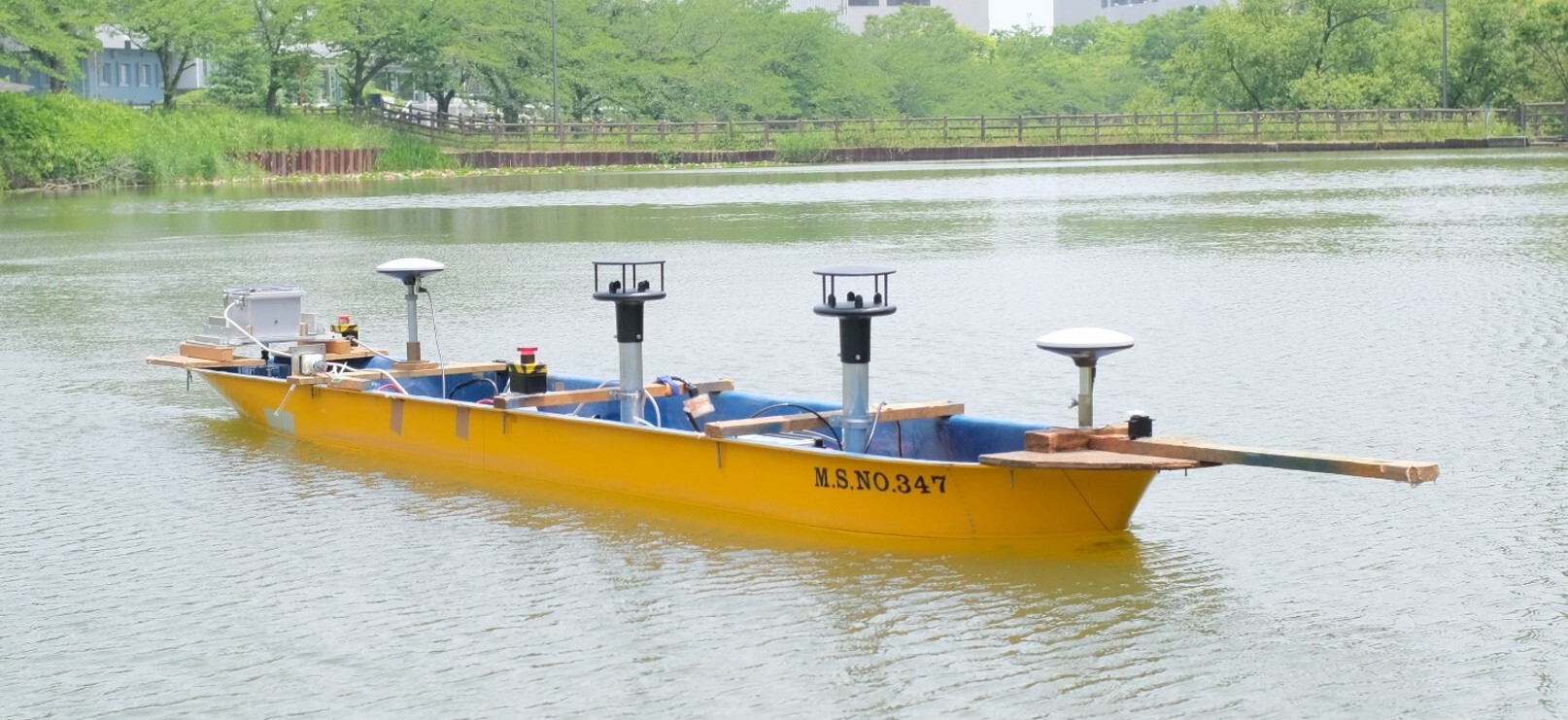}
            \caption{Photograph of the subject ship.}
            \label{fig:esso}
        \end{figure}

        The subject ship was a model ship of \textit{M.V. ESSO OSAKA} (\cref{fig:esso}).
        A nonlinear maneuvering model \cref{eq:Norrbin} was adopted in the numerical experiments.
        The parameters of the maneuvering model \cref{eq:Norrbin}, the limits on the constraints \cref{eq:magnitude_saturation,eq:rate_saturation} are summarized in \cref{tab:param_setting,tab:sat_setting}, respectively.
        \begin{table}[tb]
            \centering
            \caption{Parameters of maneuvering model used in the numerical simulation.}
            \begin{tabular}{c|cccccc}
                \hline
                Item & $K$ & $T$ & $n_{0}$ & $n_{1}$ & $n_{2}$ & $n_{3}$  \\
                \hline
                Value & $0.21$ & $8.8$ & $0$ & $0.41$ & $0$ & $0.23$  \\
                \hline
            \end{tabular}
            \label{tab:param_setting}
        \end{table}
        \begin{table}[tb]
            \centering
            \caption{Threshold values of constraints (\cref{eq:magnitude_saturation,eq:rate_saturation}) considered in the numerical simulation.}
            \begin{tabular}{c|cc}
                \hline
                Item & $M ~ \mathrm{[deg]}$ & $R ~ \mathrm{[deg/s]}$  \\
                \hline
                Value & $35$ & $20$  \\
                \hline
            \end{tabular}
            \label{tab:sat_setting}
        \end{table}
        The parameters in \cref{tab:param_setting} are determined by the system identification method using time series data of the free-running tests of the subject ship.
        The limits on the constraints; $M$ and $R$ were determined based on the mechanical constraints of the subject ship.
        The design parameters in the derived cascade system \cref{eq:state_eqs} and in the controller $\alphaeta$ are chosen as $\kdelta = \kxi = 1$ and $c_{1} = c_{2} = c_{3} = c_{4} = 1$.
        The time series were calculated by the Euler method for Case1, and by the Euler-Maruyama method for Case2.
        In all numerical simulations, the time width $\Delta t = 0.01 ~ \mathrm{s}$ was set.
        Initial states were set as $\psi(0) = r(0) = \delta(0) = \xi(0) = 0$ in Case1 and 2.

        The proposed steering control was applied for these two cases.


        \subsubsection{Case1: Heading tracking}

            In Case1, the proposed steering control was applied for heading tracking control.
            The proposed control law is designed for the tracking control with input magnitude and rate which freely behave within the constraints.
            However, with the current techniques of the authors, the computational problem with numerical saturation in the proposed method, which is detailed in \cref{sec:discussion}, can not be solved.
            Therefore, in this case, the following smooth function was adopted as the target signal.
            \begin{equation}
                \label{eq:psid_case1}
                \psid(t) = \frac{1}{2} \Psi^{\mathrm{d}} \Big( 1 + \tanh \frac{t - t_{\tanh}}{ d_{\tanh} } \Big)
                \enspace ,
            \end{equation}
            where $t_{\tanh} = 5 + 0.3 \Psi^{\mathrm{d}}$, $d_{\tanh} = 2.5 + 0.15 \Psi^{\mathrm{d}}$, and $\Psi^{\mathrm{d}}$ is the value of $\psid(t)$ at $t \rightarrow \infty$.
            Five scenarios with $\Psi^{\mathrm{d}} = 10, 20, 30, 40, 50$ were simulated.
            

        \subsubsection{Case2: Course keeping under disturbance}

            In Case2, the proposed steering control was applied for course keeping control under stochastic disturbance to check the robustness of the proposed method.
            The reference signal was set as $\psid(t) = 0$.
            In Case2, the following system haveing the form of stochastic differential equation (SDE) was considered:
            \begin{equation}
                \label{eq:dynamics_stochastic}
                \mathrm{d} r(t) = \big( f(r(t)) + b \delta(t) \big) \mathrm{d} t + \sigma \mathrm{d} W(t)
                \enspace ,
            \end{equation}
            where the Weiner process was introduced as additive noise to the model \cref{eq:dynamics} with $\sigma > 0$.
            Therefore, the inclusion of Wong-Zakai correction term is not necessary.
            This noise can be considered as a modeling error, external disturbance such as wind, or observation noise.
            In this study, we set $\sigma = b M = 0.835$, which is equivalent to the maximum influence of rudder force on the $\dot{r}$.
            \cref{eq:dynamics_stochastic} was numerically solved by the Euler-Maruyama method:
            \begin{equation}
                \begin{aligned}
                    r(t + \Delta t)
                        &= r(t) + \big( f(r) + b \delta(t) \big) \Delta t  \\
                        & \quad + \sigma ( W(t + \Delta t) - W(t) ) \enspace ,
                \end{aligned}
            \end{equation}
            where $( W(t + \Delta t) - W(t) )$ follows the normal distribution:
            \begin{equation}
                \mathcal{N}(0, \Delta t) = \sqrt{\Delta t} \mathcal{N}(0, 1)
                \enspace .
            \end{equation}

    \subsection{Result}

        \subsubsection{Case1: Heading tracking}

\begin{figure*}[tb]
                \centering
                \includegraphics[width=1.0\hsize]{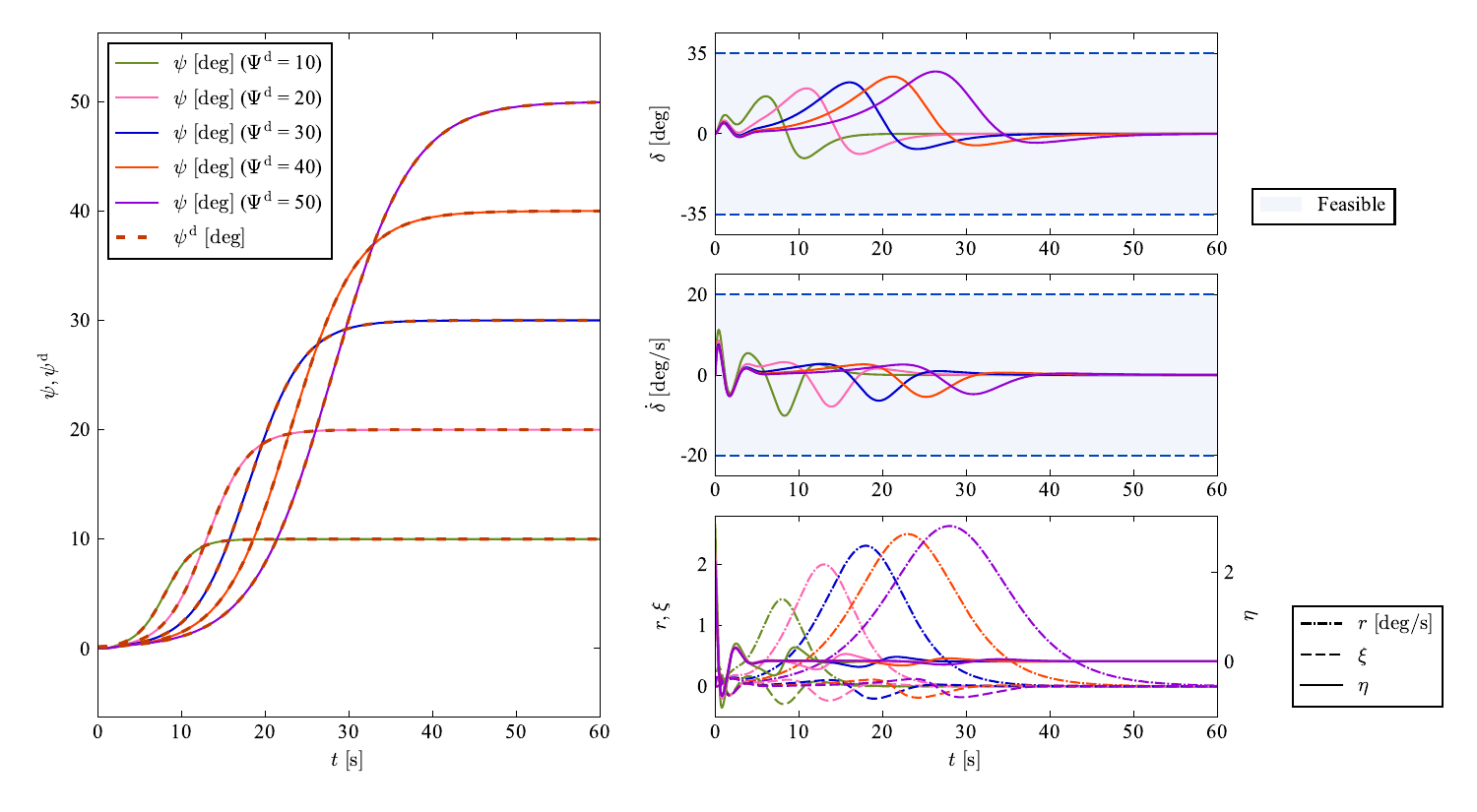}
                \caption{
                    Case1.
                    The proposed control law was applied for heading tracking control.
                    The target signal $\psid(t)$ was \cref{eq:psid_case1}.
                }
                \label{fig:case1}
            \end{figure*}

            The time series simulated in Case1 is shown in \cref{fig:case1}.
            In Case1, the proposed steering control law was applied for heading tracking control where the target signal is formulated as \cref{eq:psid_case1}.
            From \cref{fig:case1}, it is confirmed that, for every case of heading change angle $\Psi^{\mathrm{d}}$, both signals of $\delta$ and $\dot{\delta}$ did not break the constraints \cref{eq:magnitude_saturation,eq:rate_saturation}, and the heading angle $\psi$ successfully tracked the target signal $\psid$.
            This result verifies the performance of the proposed steering control law for a mild target signal.
    


        \subsubsection{Case2: Course keeping under disturbance}

\begin{figure*}[tb]
                \centering
                \includegraphics[width=1.0\hsize]{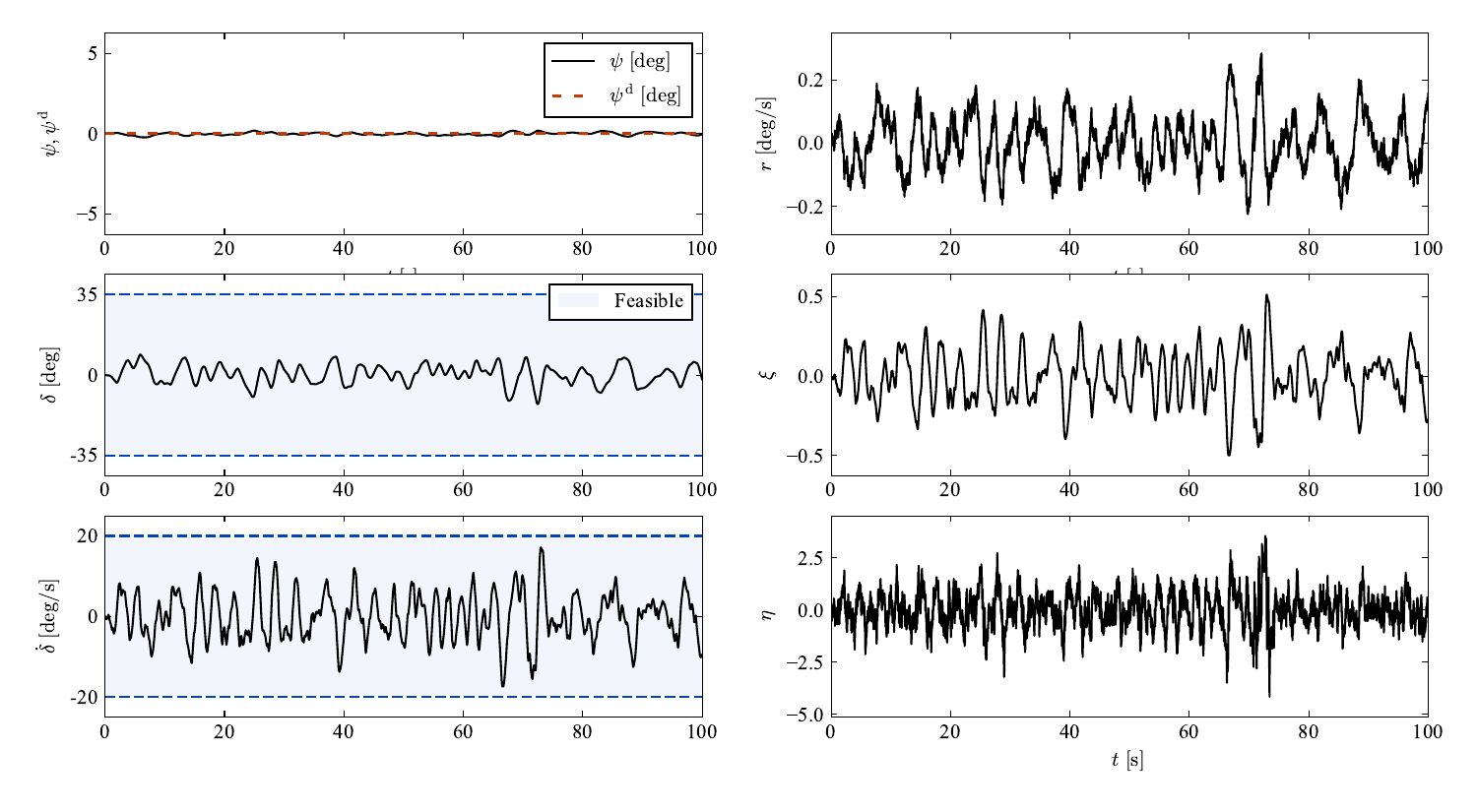}
                \caption{
                    Case2.
                    The proposed control law was implemented in the course keeping control with external disturbance.
                }
                \label{fig:case2}
            \end{figure*}

            The time series simulated in Case2 is shown in \cref{fig:case2}.
            In Case2, the proposed steering control was applied for course keeping control under stochastic disturbance.
            Stochastic noise can be observed in the time series of $r$.
            Even with this stochastic noise, the course deviation was successfully controlled with the proposed method within the input magnitude and rate constraints.
            This result shows the robustness of the proposed method to an stochastic noise to some extent.

\section{Discussion and limitation}
    
    \label{sec:discussion}

    The proposed steering control can achieve heading tracking with exponential convergence of tracking error under the constraints of rudder angle and steering speed, as shown in \cref{sec:stability}.
    Theoretically, the proposed method enables the tracking control that makes full use of almost all feasible magnitude and rate of rudder manipulation.
    In addition, due to the formulation \cref{eq:u_tanh,eq:xi_tanh}, the auxiliary system introduced by the authors has a mechanism to avoid saturation of input magnitude and rate, and $\delta$ and $\dot{\delta}$ never reach the thresholds of constraints.

    However, the proposed method has drawbacks in terms of numerical implementation.
    The cascade system \cref{eq:state_eqs} and the controller $\alphaeta$ are valid as long as the states $\delta$ and $\dot{\delta}$ are not too close to the thresholds of the constraints \cref{eq:magnitude_saturation,eq:rate_saturation}.
    However, the authors found that in the case these states get too close to the thresholds, effective solutions are unavailable.
    This is because the proposed control method does not ensure the boundedness of all signals in the closed loop.
    For example, in the third equation of \cref{eq:state_eqs}, as $\xi$ that makes $|\delta|$ approaches $M$ continues to be input, the value of $\gdelta(\delta)$ approaches zero.
    This leads to the divergence of the right hand side of \cref{eq:state_eqs} and the output of the controller \cref{eq:alpha_eta}.
    As a result, due to numerical overflow, a time series cannot be obtained unless the time width is infinitely small.
    Such control would be performed in the case a large rudder angle or/and rapid manipulation of the rudder is required, such as a large angle change of heading.
    This problem can be avoided to some extent by tuning design parameters $c_{i} ~ (i = 1, 2, 3, 4)$.
    At the present stage, it is better to shape a smooth reference signal for course change control, as exemplified in Case1 (\cref{fig:case1}).
    Future work includes improving the design of control law and numerical processing to obtain a steering control that overcomes this limitation.

\section{Conclusion}
\label{sec:conclusion}

    A ship steering control for a nonlinear system with constraints of both input magnitude and rate is proposed.
    The satisfaction of all input constraints is guaranteed by introducing a bounded smooth $\tanh$ function and auxiliary variables.
    Furthermore, using the feature of the derivative of $\tanh$ function, the time derivatives of the newly formulated state variables are calculated without auxiliary variables, and a strict-feedback system without any input constraints is derived.
    The proposed control law is designed based on the backstepping method, and the local exponential stability of the tracking error is proven.
    In the numerical experiments, it is shown that the proposed control law successfully avoids saturation of input magnitude and rate and achieves the tracking of the target heading angle.
    The unboundedness of the auxiliary systems and the constructed control law limit the proposed method, and these problems will be treated in future studies.

\begin{acknowledgements}

This study was supported by a Grant-in-Aid for Scientific Research from the Japan Society for Promotion of Science (JSPS KAKENHI Grant \#22H01701).
The study also received assistance from the Fundamental Research Developing Association for Shipbuilding and Offshore (REDAS) in Japan (REDAS23-5(18A)).
Finally, the authors would like to thank Prof. Naoya Umeda, Asst. Prof. Masahiro Sakai, Osaka University, and Prof. Hiroyuki Kajirawa, Kyushu University, for technical discussion.

\end{acknowledgements}

%
\section*{Conflict of interest}

The authors declare that they have no conflict of interest.





\bibliographystyle{spphys}       
\bibliography{main_arxiv}   

\end{document}